\documentclass[prl,twocolumn,showpacs]{revtex4-1}
\usepackage[latin9]{inputenc}
\usepackage{amsmath}
\usepackage{graphicx}
\usepackage{esint}

\usepackage[T1]{fontenc}
\usepackage{units}
\usepackage{amsbsy}
\usepackage{amssymb}
\usepackage{xargs}[2008/03/08]

\usepackage{color}

\makeatletter

\@ifundefined{textcolor}{}
{%
 \definecolor{BLACK}{gray}{0}
 \definecolor{WHITE}{gray}{1}
 \definecolor{RED}{rgb}{1,0,0}
 \definecolor{GREEN}{rgb}{0,1,0}
 \definecolor{BLUE}{rgb}{0,0,1}
 \definecolor{CYAN}{cmyk}{1,0,0,0}
 \definecolor{MAGENTA}{cmyk}{0,1,0,0}
 \definecolor{YELLOW}{cmyk}{0,0,1,0}
}

\usepackage{bbm}

\newcommandx\ket[1][usedefault, addprefix=\global, 1=]{|#1\rangle}

\newcommandx\avg[1][usedefault, addprefix=\global, 1=]{\langle#1\rangle}

\newcommandx\var[1][usedefault, addprefix=\global, 1=]{\langle(\Delta#1)^{2}\rangle}

\global\long\def\mathclap#1{\text{\hbox to 0pt{\hss\ensuremath{\mathsurround=0pt#1}\hss}}}

\begin{document}

\title{High-Capacity Angularly-Multiplexed Holographic Memory\\
 Operating at the Single Photon Level}

\author{Rados{\l}aw Chrapkiewicz}

\affiliation{Institute of Experimental Physics, Faculty of Physics, University of Warsaw, Pasteura 5, 02-093 Warsaw, Poland}

\author{Micha{\l} D\k{a}browski}

\email{mdabrowski@fuw.edu.pl}

\affiliation{Institute of Experimental Physics, Faculty of Physics, University of Warsaw, Pasteura 5, 02-093 Warsaw, Poland}

\author{Wojciech Wasilewski}

\affiliation{Institute of Experimental Physics, Faculty of Physics, University of Warsaw, Pasteura 5, 02-093 Warsaw, Poland}

\date{\today}


\begin{abstract}
We experimentally demonstrate an angularly-multiplexed holographic
memory capable of intrinsic generation, storage and retrieval of multiple photons,
based on off-resonant Raman interaction in warm $\mbox{rubidium-87}$
vapors. The memory capacity of up to 60 independent atomic spin-wave
modes is evidenced by analyzing angular distributions of coincidences
between Stokes and time-delayed anti-Stokes light, observed down
to the level of single spin-wave excitation during several-$\mu$s memory lifetime. We also propose how to practically
enhance rates of single and multiple photons generation by combining
our multimode emissive memory with existing fast optical switches.
\end{abstract}
\maketitle
Construction of on-demand sources of desired quantum states of light
remains an overarching goal for quantum engineering. While single
photons are essential resource for quantum communication protocols
\citep{Duan2001,Yuan2008}, multiple photon states offer an avenue for quantum
computing with linear optics \citep{Knill2001,Kok2007} and quantum
simulations, e.g. in boson sampling schemes \citep{Peruzzo2010,Motes2014}.
Since scientists are already able to manufacture complex photonic
chips using femtosecond writing technologies \citep{Gross2015}, perhaps
the last major roadblock to perform linear-optics simulations unattainable
by classical computing is the ability to create large number of photons
capable of non-classical interference~\citep{Jachura2014}.

Prospective solutions to achieve this non-trivial, long-standing goal
rely on still developing quantum dot sources \citep{Gazzano2013,Ding2016,Bennett2016}, Rydberg blockade \citep{Peyronel2012,Ripka2016}
as well as parametric processes such as four-wave mixing \citep{Alibart2006a}
and spontaneous parametric down-conversion (SPDC) routinely employed
to produce heralded single photons \citep{URen2004}. The present
technology of SPDC sources is well developed and widespread since,
while operating in room temperatures, it offers high brightness and
renders low noise. Nonetheless, the intrinsic feature of parametric
sources is their purely stochastic operation. The probability of photon
pairs generation must be kept low to suppress the contribution of
higher numbers of photons. A possible way to surpass this stochastic
behavior is to combine multiple sources with fast, active optical
switches \citep{Hall2011,Ma2011} to increase chances for single photon
generation. In practice, setups with at most four sources using SPDC \citep{Yao2012,Collins2013} and twelve sources with cold atoms \citep{Lan2009} have been demonstrated. Furthermore, $N$ parametric
sources can be used to generate $N$ heralded photons but this method
requires very long waiting time \citep{Yao2012} as the probability
for $N$-photon state generation falls exponentially with $N$.

\begin{figure*}[t]
\includegraphics[width=1\textwidth]{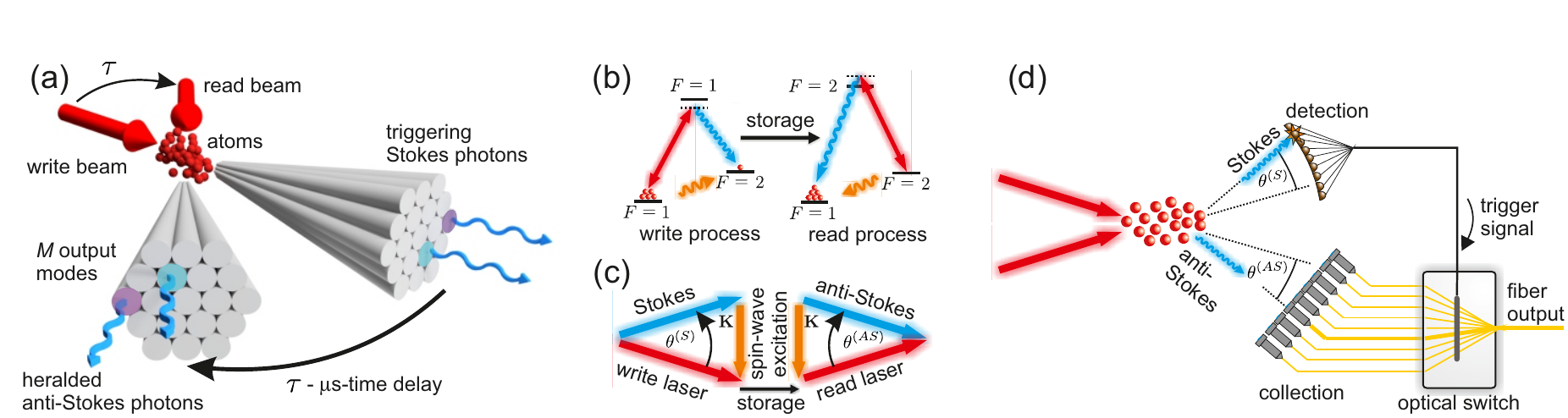}\protect\caption{(a) Idea of holographic memory generating, storing and releasing on-demand
angularly correlated photons. Photons are produced in $2\times M$
modes which are pairwise coupled. In each independent pair of modes
(e.g. colored circles), anti-Stokes photons are heralded by Stokes
photons generation, with $\mu$s-time delay advance. (b) Generation
and retrieval of photons is performed in Raman scattering process,
using $\Lambda$-system $\mathrm{^{87}Rb}$ energy levels. (c) Phase-matching
condition relates wave-vectors of driving laser beams, photons and
spin-wave excitation and determines the scattering angle of an anti-Stokes
photon $\boldsymbol{\theta}^{\mathrm{(AS)}}\simeq-\boldsymbol{\theta}^{\mathrm{(S)}}$.
(d) Predictability of scattering angles of anti-Stokes photons can
be utilized to match outgoing photons to optical switch \citep{Hall2011,Ma2011}
with one or multiple outputs (not shown). Switch architecture is reconfigured
after detection of the Stokes trigger photons so as to pass single
or $N$-photons into desired fiber links. \label{fig:Idea-router}}
\end{figure*}

Recently, Nunn \emph{et al.} \citep{Nunn2013} have suggested a possible
solution for enhanced generation of $N$-photon states in a system
with $N$ quantum memories storing heralded photons from $N$ independent
SPDC sources, and releasing them at once. Here we present a different
approach where photons can be generated directly inside many emissive
quantum memories driven by Raman scattering process where deposition
of an excitation is heralded by a detection of a Stokes photon. Moreover,
as depicted in Fig.~\ref{fig:Idea-router}(a-c), multiple memories
are implemented in our experiment within one atomic ensemble which
serves as a single holographic memory interfacing many overlapping
spin-wave modes  of different spatial periodicity with independent
angular modes of light. Presented angular multiplexing of many modes
provides a tremendous simplification of experimental setups as compared
with hypothetical multiple stand-alone memory systems.

Similarly as in SPDC sources the probability $\zeta$ of generating
single excitation in any of memory modes has to be low, to suppress
the probability $\zeta^{2}$ for generating two excitations. However,
if the number of available emissive modes $M$ is large, the probability
of photon generation in at least one of them $\mathrm{1-(1-\zeta)}^{M}$ 
 can approach unity \citep{Ma2011}. For instance for $\zeta=10^{-2}$,
$M\mathrm{=100}$ the probability for Stokes photon generation reaches
60\%. Once a photon is emitted in any of $M$ modes, one can use an
active optical switch \citep{Hall2011,Ma2011} seeded by triggering
signal from the Stokes scattering, to route the anti-Stokes photon
to a desired single output, as illustrated in Fig.~\ref{fig:Idea-router}(d).
Such operation essentially relies on a memory storage time which has
to exceed nanoseconds-long reconfiguration time of the optical switch
\citep{Hall2011,Ma2011}.

Note that probability for $M$ modes to generate exactly $N$ photons
${M \choose N}\zeta^{N}(1-\zeta)^{M-N}$ can be also high. For $N=8$,
$M=100$ it reaches a value of $7\times10^{-6}$ which is over ten
orders of magnitude larger than probability to obtain 8 photons from
eight independent sources $\zeta^{8}=10^{-16}$. The scheme of Fig.~\ref{fig:Idea-router}(d)
could be upgraded with a  multiple-outputs switch. This way, after each
successful generation of $N$ excitations, the same multiphoton state
of the product form $\ket[1]\otimes\ldots\otimes\ket[1]$ can be produced
in a desired set of output fibers. Most of the elements of this scheme
such as matrices of single photon detectors, fiber bundles and fast
optical switches have been demonstrated or are commercially available.
The centerpiece is the multiplexed source of heralded photons
with sufficient time-delay, based on spatial \citep{Lan2009}, temporal \citep{Simon2010, Albrecht2015} or another degree of freedom.

In this Letter we experimentally demonstrate a holographic atomic
memory that can generate, store and retrieve light at the single photon level in many independent
angular modes whose number could practically reach thousands \citep{Grodecka-Grad2012}
under realistic experimental conditions. Rather than using external
non-deterministic sources of spectrally-filtered, heralded photons
to convert them to atomic collective excitations \citep{Michelberger2014},
the Stokes photon generated inside the memory heralds the spin-wave
creation which can be retrieved on demand in the time-delayed anti-Stokes
scattering process. Combination of large number of emissive modes
with predictability offered by a memory time-delay can circumvent
the limits of SPDC sources and serve as an enhanced-rate source of
photons. Moreover, our approach bridges the gap between numerous experiments
with warm atomic memories with single spatial mode of light at the
single photon level \citep{VanderWal2003,Reim2011,Bashkansky2012,Michelberger2014}
and a few spatially multimode experiments performed at macroscopic
light levels \citep{Higginbottom2012,Firstenberg2013a}. 

Operation of our memory presented in Fig.~\ref{fig:Idea-router}(a-c)
relies on collective multimode Raman scattering in atomic vapors in
a DLCZ scheme \citep{Duan2001}. The angular correlations between
Stokes and anti-Stokes photons, mediated by phase-matching conditions
depicted in Fig.~\ref{fig:Idea-router}(c), arise in a way similar
to holography \citep{Dai2012}. Firstly we write to the memory by
driving spontaneous Raman transition in $\Lambda$-system, producing
pairs of collective atomic excitation and spontaneous Stokes photon
scattered in random direction. The write process resembles registering
of a hologram and depending on the Stokes photon scattering angle
$\boldsymbol{\theta}^{\mathrm{(S)}}$ the spin-wave with wave-vector
$\mathrm{\mathbf{K}}=2\pi\boldsymbol{\theta}^{\mathrm{(S)}}/\lambda$
is created, where $\lambda$ stands for a wavelength of a driving
beam. Then, after adjustable storage time $\tau$, the atomic excitation
can be converted to anti-Stokes photon with up to $\eta_{\mathrm{read}}=30\%$ internal read efficiency \citep{Reim2011,Dabrowski2014,Supplemental}. The read
process is analogous to reconstruction of a hologram and the angle
of the anti-Stokes photon emission $\mathbf{\boldsymbol{\theta}^{\mathrm{(AS)}}\simeq-\boldsymbol{\theta}^{\mathrm{(S)}}}$
is determined by the preceding detection of the Stokes photon direction,
a few microseconds in advance. Remarkably, the holographic storage
allows us to populate, store and retrieve multiple overlapping yet
independent spin-wave modes in the same ensemble of atoms.

\begin{figure}[b]
\includegraphics[width=8.6cm]{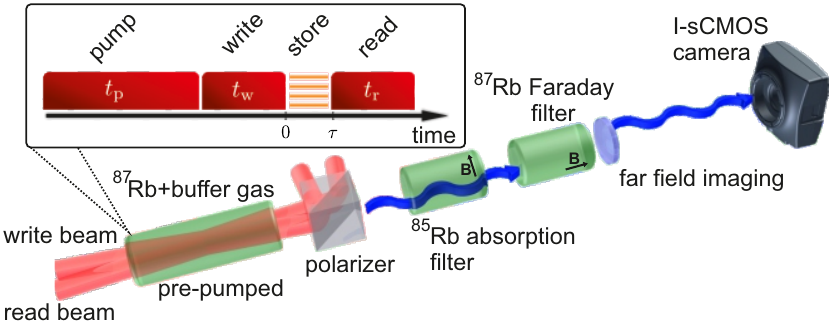}\centering 

\protect\caption{The simplified scheme presenting key parts of the experimental setup
for the holographic memory. Generation and storage of photons are
implemented in warm $\mathrm{^{87}Rb}$ vapors whose thermal motion
is slowed down by admixture of noble buffer gas. Inset: experimental
sequence. The triple filtering system \citep{Dabrowski2015} (polarization
and two-step spectral, $\mathbf{B}$ denotes magnetic field vector)
passes scattered photons at different angles and blocks all laser
beams detuned by 6.8~GHz. Detection of photons with high angular
resolution is performed with an intensified sCMOS camera (I-sCMOS)
capable of resolving photon coincidences \citep{Jachura2015,Supplemental}.\label{fig:Experimental-setup}}
\end{figure}

The simplified scheme of the experimental setup is depicted in Fig.~\ref{fig:Experimental-setup}
(cf. \citep{Chrapkiewicz2012,Dabrowski2015} for details). We pump atoms into $\mathrm{5^{2}S_{1/2}},\,\mathrm{F=1}$ and drive
Raman transition to $\mathrm{5^{2}S_{1/2}},\,\mathrm{F=2}$ level.
The driving lasers detuning was approximately 700~MHz from $\mathrm{5^{2}P_{1/2}},\,\mathrm{F=1}$ or $\mathrm{5^{2}P_{3/2}},\,\mathrm{F=2}$ levels as marked
in Fig.~\ref{fig:Idea-router}(b). The main glass cell of $L=10$~cm
length was filled with $\mathrm{^{87}Rb}$ mixed with krypton at 1
Torr and heated to $\mathrm{60^{\circ}-75^{\circ}}$~C corresponding
to optical depths (OD) from 50 up to 80. The $1/e^{2}$ diameters
of the linearly polarized driving beams at 795~nm or 780~nm tilted by 10 mrad inside the cell were about
$2w=4.6$~mm while the pump beam at 780~nm was twice as large. The
typical experimental sequence depicted in the inset of Fig.~\ref{fig:Experimental-setup}
consists of pulses of 70~mW pump, 20~mW write and 15~mW read laser
of duration: $t\mathrm{_{p}\approx1\ ms}$, $t_\mathrm{w}\simeq t_{\mathrm{r}}\approx1\,\mu s$
while storage period is varied in the range $0\leq\tau\leq40\,\mu\mathrm{s}$.
The triple filtering system --- Wollaston prism, atomic absorption
filter and Faraday filter (cf. \citep{Dabrowski2015} for details)
is used to perform coincidence measurements at the single photon level.
Essential adventage of our solution over routine spectral filtering
using Fabry-Perot interferometers, is its transmission being insensitive
to a photon scattering angle. We obtain large transmission of ca.
50\% for both Stokes and anti-Stokes photons while attenuating 6.8~GHz
detuned driving beams by the factor of $10^{11}$. Moreover narrow
spectral windows of the Faraday filter suppress the broadband collisional
fluorescence \citep{VanderWal2003} by at least two orders of magnitude
and virtually blocks contribution from the four-wave mixing during read process \citep{Dabrowski2014}. 
Finally, we split Stokes from anti-Stokes using interference filters and angularly resolve them with suitable focusing lenses. 
Each field is directed to a separate regions on the  intensified scientific CMOS (I-sCMOS) camera \citep{Chrapkiewicz2014a,Jachura2015,Supplemental} 
used for counting coincidences.
\begin{figure}[t]
\includegraphics[width=1\columnwidth]{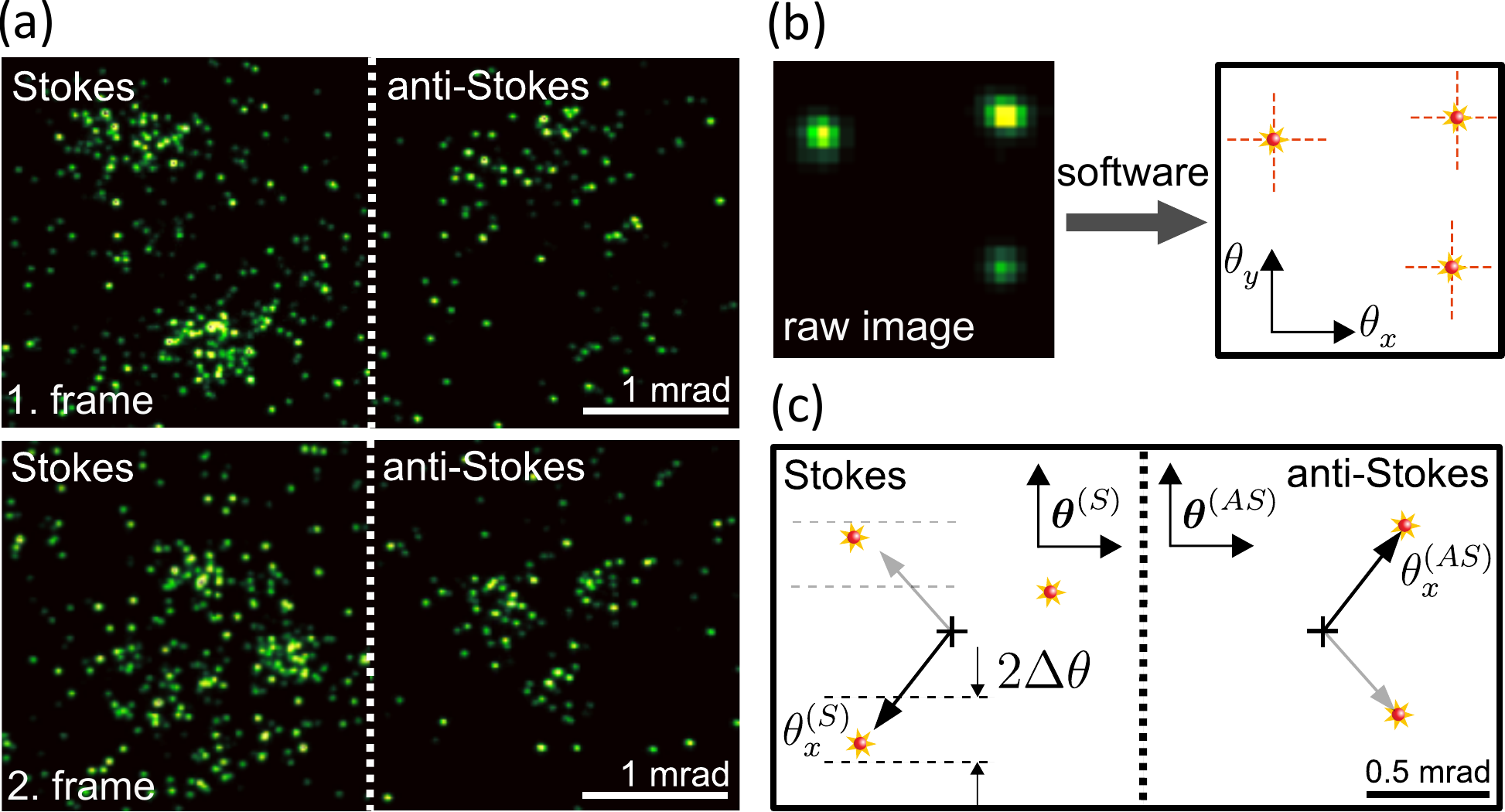}\centering

\protect\caption{(a) Two exemplary cropped frames with Stokes and 500~ns-delayed anti-Stokes
photons detected as bright spots on I-sCMOS camera. Hundreds of angularly-resolved
photons from a high-gain Raman scattering form pairwise inverted random
patterns. (b) Real-time software algorithm converts with sub-pixel
resolution raw I-sCMOS images into sets of coordinates representing
central positions of individual single photon spots \citep{Jachura2015}.
(c) Exemplary processed frame with several photons scattered in a
low-gain regime. We further accumulate histograms of photon coordinates
$n_{\mathrm{coinc}}(\theta_{x}^{\mathrm{(S)}},\theta_{x}^{(\mathrm{AS})})$
for coincidences where detected anti-Stokes photon was preceded by
a Stokes trigger found in a stripe fulfilling phase-matching condition
$\theta_{y}^{(\mathrm{S})}=-\theta_{y}^{(\mathrm{AS})}\pm\Delta\theta$,
where $\Delta\theta=0.15$ mrad or 0.3 mrad. Black crosses mark driving
beams directions. \label{fig:Single-shots}}
\end{figure}

Exemplary, raw single-shot images of the Stokes and the 500 ns-delayed
anti-Stokes scattering are presented in Fig.~\ref{fig:Single-shots}(a).
Visibly separated spots originate from multiple Raman scattered photons
impinging the camera image intensifier where they are converted to
bright phosphor flashes. The registered photons are noticeably grouped
in speckle-like patterns \citep{Chrapkiewicz2012} which in the anti-Stokes
process are angularly-inverted as compared to the Stokes scattering,
following the phase matching condition in Fig.~1(c). Here we exemplify
an outcome of a high-gain Raman process where hundreds of photons
are scattered in each shot, for $t_{\mathrm{w}}=2\ \mathrm{\mu s}$,
$t_{\mathrm{r}}=1\ \mathrm{\mu s}$ and $\mathrm{OD}=80$. Thereafter
we decreased the number of Stokes scattered photons whose mean number
depends exponentially (high-gain) on write pulse duration $t\mathrm{_{w}}$ and
OD. From each captured frame we extract the coordinates of individual
photon detection events $\boldsymbol{\theta}^{(\mathrm{S})}=(\theta_{x}^{(\mathrm{S})},\theta_{y}^{(\mathrm{S})})$,
$\boldsymbol{\theta}^{(\mathrm{AS})}=(\theta_{x}^{(\mathrm{AS})},\theta_{y}^{(\mathrm{AS})})$
for Stokes and anti-Stokes separate camera regions, with a real-time software
processing \citep{Chrapkiewicz2014a,Jachura2015} as visualized in
Fig.~\ref{fig:Single-shots}(b). An exemplary processed frame in the low-gain regime for
$t_{\mathrm{w}}=250\ \mathrm{ns}$, $t_{\mathrm{r}}=500\ \mathrm{ns}$ and $\mathrm{OD}=50$
is shown in Fig.~\ref{fig:Single-shots}(c). 

We quantify the angular properties of the Raman scattering by measuring
and analyzing distributions of coincidences $n_{\mathrm{coinc}}(\boldsymbol{\theta}^{(\mathrm{S})},\boldsymbol{\theta}^{(\mathrm{AS})})$
between the Stokes and time-delayed anti-Stokes photons. To confirm
the angular correlations between time-delayed photons $\boldsymbol{\theta}^{(\mathrm{AS})}\simeq-\boldsymbol{\theta}^{(\mathrm{S})}$
it is convenient to display and focus on bidimensional coincidence
distributions such as $n_{\mathrm{coinc}}(\theta_{x}^{(\mathrm{S})},\theta_{x}^{(\mathrm{AS})})$,
equivalent to measurement with two one-dimensional detectors, cf.
\citep{Peeters2009}. As illustrated in Fig.~\ref{fig:Single-shots}(c),
to utilize all photons detected in two-dimensional camera frames we
take into account photon pairs that appear in conjugate $\theta_{y}$-separated
stripes such $|\theta_{y}^{(\mathrm{S})}+\theta_{y}^{(\mathrm{AS})}|<\Delta\theta$,
integrating over $\theta_{y}^{(\mathrm{S})},\theta_{y}^{(\mathrm{AS})}$
coordinates: 
\begin{equation}
n_{2}(\theta_{x}^{(\mathrm{S})},\theta_{x}^{(\mathrm{AS})})=\int d\theta_{y}^{(\mathrm{AS})}\int\limits _{\mathclap{\theta_{y}^{(\mathrm{AS})}-\Delta\theta}}^{\mathclap{\theta_{y}^{(\mathrm{AS})}+\Delta\theta}}d\theta_{y}^{(\mathrm{S})}\,n_{2}(\boldsymbol{\theta}^{(\mathrm{S})},\boldsymbol{\theta}^{(\mathrm{AS})}).
\label{Eq:n_2}
\end{equation}
The total number of coincidences counted this way
$n_{2}(\theta_{x}^{(\mathrm{S})},\theta_{x}^{(\mathrm{AS})})$  
is a sum of two factors: 
a total number of Stokes--anti-Stokes pairs 
$n_{\mathrm{coinc}}(\theta_{x}^{(\mathrm{S})},\theta_{x}^{(A\mathrm{S})})$
and accidental coincidences from uncorrelated noise 
$n_{\mathrm{acc}}(\theta_{x}^{(\mathrm{S})},\theta_{x}^{(\mathrm{AS})})$
whose spatial distribution reads as follows:
\begin{equation}
n_{\mathrm{acc}}(\theta_{x}^{(\mathrm{S})},\theta_{x}^{(\mathrm{AS})})=
\frac{1}{\mathcal{N}}\int d\theta_{y}^{(\mathrm{AS})}
\int\limits _{\mathclap{\theta_{y}^{(\mathrm{AS})}-\Delta\theta}}^{\mathclap{\theta_{y}^{(\mathrm{AS})}+\Delta\theta}}
d\theta_{y}^{(\mathrm{S})}\,n_{1}(\boldsymbol{\theta}^{(\mathrm{S})})
\,n_{1}(\boldsymbol{\theta}^{(\mathrm{AS})}),
\label{Eq:n_acc}
\end{equation}
where $n_{1}(\boldsymbol{\theta}^{(S)}),n_{1}(\boldsymbol{\theta}^{(AS)})$
are distributions of number of photons collected in both camera regions
and $\mathcal{N}$ is the total number of frames. 
Therefore the number of Stokes--anti-Stokes pairs can be calculated by subtracting the Eq.~(\ref{Eq:n_acc}) from Eq.~(\ref{Eq:n_2}) \citep{Peeters2009}.
At the center of the distribution the accidential coincidendes constitute about 58\% of the total number of pairs.

\begin{figure}[t]
\includegraphics[width=1\columnwidth]{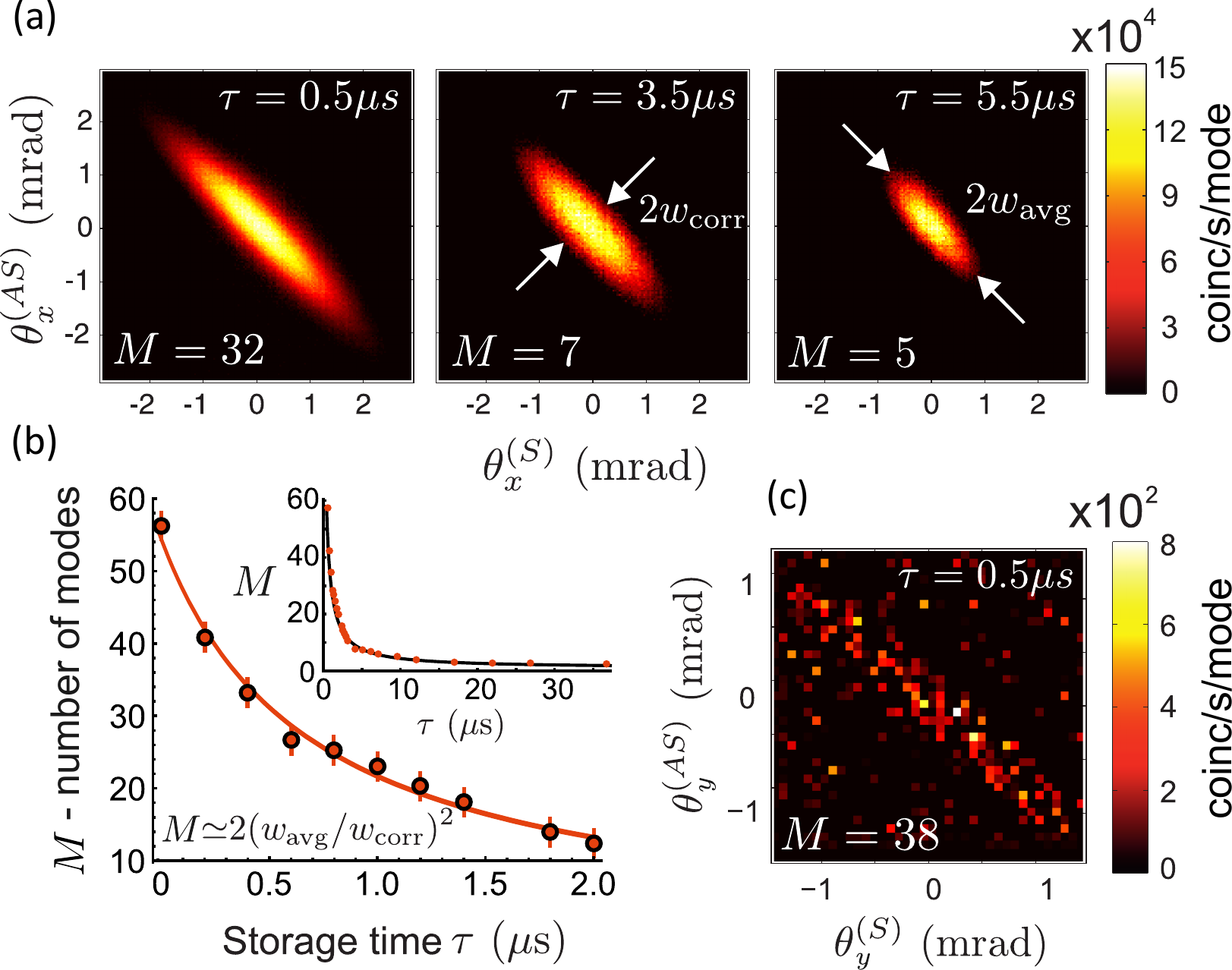}\centering

\protect\caption{(a) Angular distributions of Stokes and anti-Stokes photon coincidences
$n_{\mathrm{coinc}}(\theta_{x}^{(\mathrm{S})},\theta_{x}^{(\mathrm{AS})})$
for increasing storage times $\tau$ reveal angular correlations $\theta_{x}^{(\mathrm{AS})}\simeq-\theta_{x}^{(\mathrm{S})}$
between photons stored and retrieved from the holographic memory.
A visible decay of high-scattering-angle events is determined by a
diffusional decoherence \citep{Chrapkiewicz2014}. (b) A doubled squared
ratio between anti-diagonal and diagonal width of coincidence histograms
is used to estimate the total number of retrieved angular modes $\mathrm{M<60\simeq\mathcal{F}}$
versus the storage time in agreement with diffusion model \citep{Supplemental}. Tens of modes are retrieved within first
$2\mathrm{\:\mu s}$, which is sufficiently long to couple the holographic
memory source with currently existing optical switches \citep{Hall2011,Ma2011}
(cf. Fig. 1(d)). Inset: a few modes can be retrieved up to tens of
microseconds. (c) Coincidences are observable down to the single spin-wave
excitation level. \label{fig:Cross-correlation-maps}}
\end{figure}

In Fig.~\ref{fig:Cross-correlation-maps}(a) we present histograms
representing angular distribution of coincidences between Stokes and
anti-Stokes photons $n_{\mathrm{coinc}}(\theta_{x}^{(\mathrm{S})},\theta_{x}^{(\mathrm{AS})})$
for three different storage times $\tau=0.5,3.5,5.5\ \mu\mathrm{s}$,
each obtained from $3\times10^{5}$ frames for $t_{\mathrm{w}}=1\ \mu s$,
$t_{\mathrm{r}}=1\ \mu s$ and $\mathrm{OD}=80$, for high Raman gain ($\eta_{\mathrm{read}}=30\%$).
The maximum number of Stokes--anti-Stokes generated pairs per mode
per single shot translates here to about 40 after correcting for the setup transmission and the detection efficiency. Characteristic elongated anti-diagonal shapes attest
the conjugation of opposite angles of Stokes and time-delayed anti-Stokes
scattering resulting from the phase matching condition $\theta_{x}^{(\mathrm{AS})}\simeq-\theta_{x}^{(\mathrm{S})}$.
The diagonal $1/e^{2}$ width of $2w_{\mathrm{corr}}=1.2$~mrad,
almost independent of a storage time, represents the angular spread of the
mode of anti-Stokes emission conditioned on preceeding Stokes localization. The anti-diagonal width $2w_{\mathrm{avg}}$
represents a span of angles $\theta_{x}{}^{(\mathrm{AS})}$ where
the anti-Stokes light is detected. It decreases during the storage
due to diffusion of atoms in the buffer gas \citep{Firstenberg2013a,Chrapkiewicz2014}
which in turn washes out dense fringes in the stored spin-hologram
coupling to higher angles of scattering \citep{Supplemental}. 

The estimated number of modes $M$ of our memory can be taken as the
ratio of the solid angle span of anti-Stokes to a single mode spread
$M\simeq2(w_{\mathrm{avg}}/w_{\mathrm{corr}})^{2}$ \citep{Chrapkiewicz2012},
taking into account full two-dimensional mode structure illustrated
in Fig.~\ref{fig:Idea-router}(a). We calculate the number of modes
$M$ on respective coincidence distributions and in Fig.~\ref{fig:Cross-correlation-maps}(b)
we present its dependence on the storage time acquired from additional
measurement series, together with curve fitting \citep{Supplemental}. $M$ dropped from 58 for an instantaneous retrieval
to \textcolor{black}{12} after $2\mathrm{\:\mu s}$ storage time.
In the inset we can observe that for the longer storage times the
number of retrieved modes drops to 2 within about $20\mathrm{\:\mu s}$.
The maximum value displayed in Fig.~\ref{fig:Cross-correlation-maps}(b)
is close to the Fresnel number of the atomic ensemble $\mathcal{F}=w^{2}/\lambda L\approx66$
as can be expected from the theory $M\simeq\mathcal{F}$ \citep{Raymer1981}.
We therefore envisage that by increasing beams diameter about 4 times
up to 18~mm, retrieval of even 1000 modes should be realistic. 

Finally, in Fig.~\ref{fig:Cross-correlation-maps}(c) we demonstrate
the operation of the memory at the single excitation level, where maximum number
of pairs generated per shot per mode is about $5\times10^{-2}$,
for $t_{\mathrm{w}}=250\ \mathrm{ns}$, $t_{\mathrm{r}}=500\ \mathrm{ns}$ and $\mathrm{OD}=50$
providing a low Raman gain ($\eta_{\mathrm{read}}=13\%$). Here coincidences distribution $n_{\mathrm{coinc}}(\theta_{y}^{(\mathrm{S})},\theta_{y}^{(\mathrm{AS})})$
acquired from $10^{6}$ frames reveals the similar structure and consequently
number of retrieved modes $M$ as in the higher gain regime. By
keeping the probability for photon pair per mode low, here we generate
on average ca. 0.9 Stokes--anti-Stokes pairs per shot in the
whole memory \citep{Supplemental}. This is an evidence of possible rate enhancement due
to the mode multiplexing, while photons are $0.5\,\mu\mathrm{s}$-delayed
in time. 

In both high and low-gain regimes we obtain the ratio of coincidences
to overall registered pairs of approx. 42\% which is comparable
with single-spatial-mode warm atomic memories \citep{Reim2011,Michelberger2014,Bashkansky2012}.
The noise of the same frequency as anti-Stokes photons originates
primarily from collisions with buffer gas \citep{Manz2007} that lead
to isotropic and incoherent Stokes scattering in the read stage.
Although an employment of even more stringent spectral filtering would
further decrease a broadband noise component as shown in a single-spatial-mode
system \citep{Bashkansky2012}, we suspect that further demonstrations
beyond proof of principle, aiming in practical $N$-photon source,
may benefit from implementations in cold atomic systems where high
non-classical correlations can be readily achieved \citep{Chaneliere2005,Bao2012,Farrera2016}. 

In conclusion, we presented the first experimental demonstration of
large number of angular modes from the emissive atomic memory that predictably generate anti-Stokes photons
in a direction known several microseconds in advance through the preceding
detection of Stokes photons. In our implementation in warm rubidium-87
vapors we were generating, storing and retrieving light at the single photon level from up
to 60 spin-wave modes whose number can be further scaled up with an
increasing Fresnel number \citep{Raymer1981,Grodecka-Grad2012}. From supporting measurements we infer that no more than one spin-wave per mode is present in the memory \citep{Supplemental}, but non-classical operation of the memory is not directly shown. We suggest that a combination of such multimode memory with existing optical
switches \citep{Hall2011,Ma2011} will be a realistic method for $\mbox{enhanced-rate}$
generation of single \citep{Matsukevich2006,Chou2007} or $N$-photon states, the latter being a prerequisite
for linear-optics quantum computing \citep{Knill2001,Kok2007} and
simulations \citep{Peruzzo2010,Motes2014}. Our work constitutes the
first observation of time-delayed multimode angular correlations at
the single photon level in contrast to all previous studies where
such correlations were observed exclusively in instantaneous nonlinear
processes \citep{Chang2014}, such as parametric down-conversion \citep{Jedrkiewicz2004a}
or four-wave mixing \citep{Boyer2008e}.

\begin{acknowledgments}\textsl{Acknowledgments.}
We acknowledge discussions with A. Bogucki,
M. Jachura, A. Leszczy\'{n}ski, M. Lipka, M. Mazelanik, M. Parniak
and P. Zi\'{n} as well as support of K. Banaszek and J. Iwaszkiewicz. This project was
financed by the National Science Center (Poland) (Grants No. DEC-2011/03/D/ST2/01941 and DEC-2015/19/N/ST2/01671),
PhoQuS@UW (Grant Agreement No. 316244) and it was supported in part
by PL-Grid Infrastructure. R.C. was supported by the Foundation for
Polish Science.
\end{acknowledgments}

\bibliographystyle{apsrev4-1}

\section*{Supplemental Material}

\section{Heralding and detection efficiency}
The efficiency of our quantum memory setup is limited by finite retrieval
efficiency of the anti-Stokes light $\eta^{\mathrm{(high)}}_{\mathrm{read}}=0.3$ for stimulated (hig-gain) regime or $\eta^{\mathrm{(low)}}_{\mathrm{read}}=0.13$ for spontaneous (low-gain) regime. The characteristic decay of signal scattered during the read process (see Fig. \ref{fig:timeEvolution}) depends on the write pulse duration and optical depth (OD). The effective retrieval time is limited by the difussion, thus in the sponteneous regime we obtain lower read efficiency than for stimulated regime. We filtered out the four-wave mixing noise \citep{Dabrowski2014S}
using Faraday filter with transmission $\eta_{\mathrm{T}}=0.5$ \citep{Dabrowski2015S}, at the same
time achieving the total heralding efficiency of $\eta^{\mathrm{(high)}}_{H}=0.15$ or $\eta^{\mathrm{(low)}}_{H}=0.07$.

This is comparable to other single-photon sources implemented using
quantum memories. Single-photon-level quantum memory using room-temperature
atomic vapors \citep{Reim2011S} as well as cold atoms \citep{Farrera2016S}
both achieve $\eta_{H}=0.2$. In experiments with bulk diamond in
room-temperature \citep{England2015S} and hydrogen molecules \citep{Bustard2013S}
picosecond storage times with heralding efficiency of $\eta_{H}=0.16$
and $\eta_{H}=0.18$ have been reported. Solid state spin-wave quantum
memory for time-bin qubits with $\eta_{H}=0.06$ has been recently
constructed \citep{Gundogan2015S}. Saunders \emph{et al}. \citep{Saunders2016S}
used external cavity to improve the ratio of signal of Raman scattered
photons to noise from the four-wave mixing process and obtained $\eta_{H}=0.1$.

Finite heralding efficiency $\eta_{H}$ limits the chance of retrieving
exactly $N$ photons (by a factor of $\eta_{H}^{N}$) to \mbox{$\eta_{H}^{N}{M \choose N}\zeta^{N}(1-\zeta)^{M-N}$}
where $\zeta$ is the probability of generating single excitation
in any of $M$ memory modes. With current heralding efficiency this
would severely limit the $N$-photon production rate. However even
this limited rate would be orders of magnitude faster for $N=8$,
$M=100$ than the rate at which 8 SPDC sources fire simultaneously
$\zeta^{N}$, even assuming they approach perfect heralding limit $\eta_{H}\simeq1$,
if the pair production rate $\zeta$ is small enough to inhibit creation
of double pairs from any source. Similarly as in SPDC sources, the source
we propose can work only with post-selection.

The detection system we use is a commercially available sCMOS Andor Zyla 5.5 Mpx camera connected with Hamamatsu Image Intensifier (V7090D). The quantum efficiency of the whole system is QE=20\%, a dark count rate of 2 counts per 1 $\mu$s image intensifier gate time for the whole camera region and with the shortest possible gating time of 100 ns.

\section{Temporal and statistical properties of scattered photons}

\begin{figure}[b]
\includegraphics[scale=0.5]{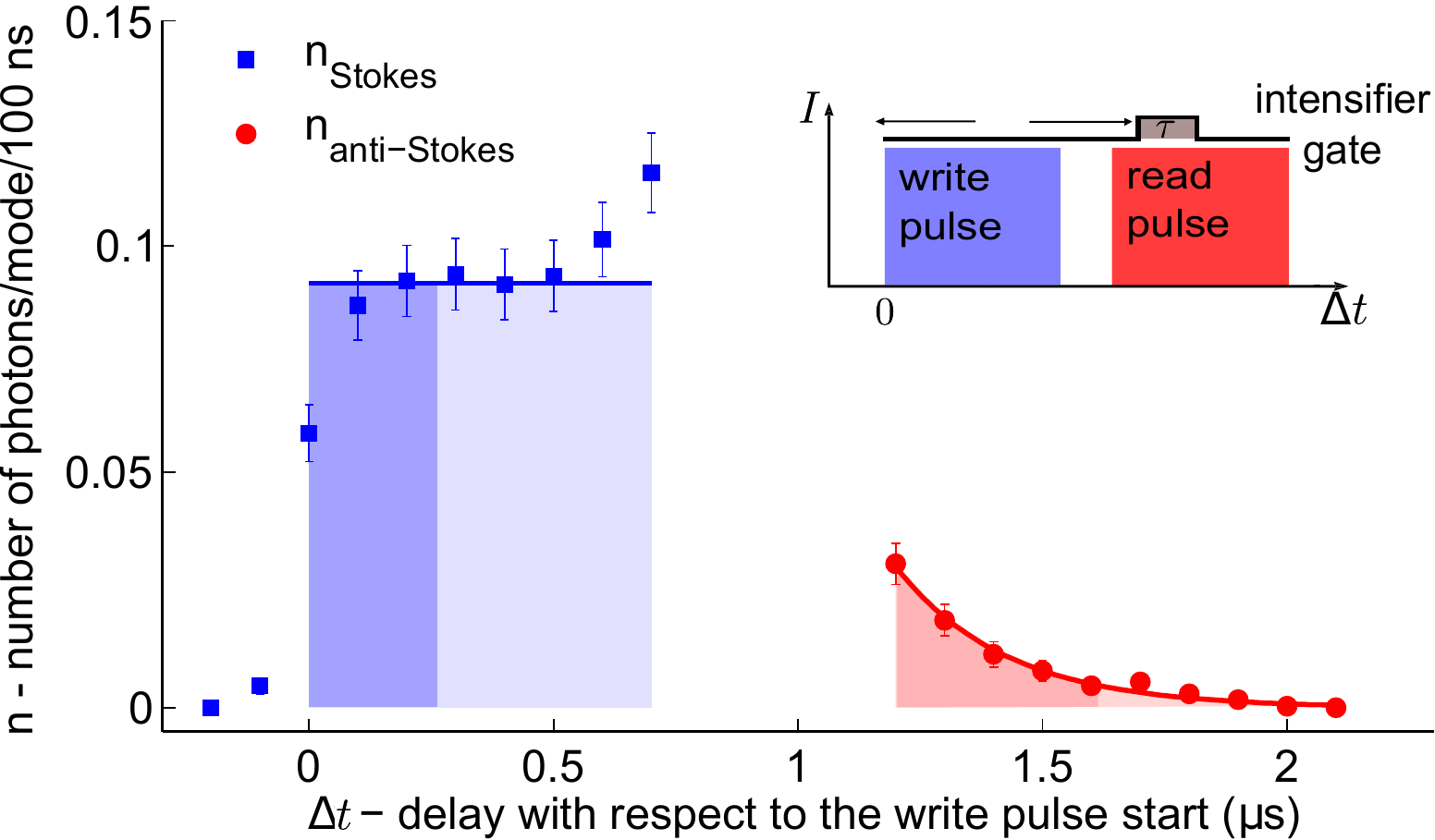}\centering \caption{Time evolution of photon number scattered per mode during 100 ns for $t_{\mathrm{w}}=700$ ns, $t_{\mathrm{store}}=500$ ns,
$t_{\mathrm{r}}=800$ ns. Stokes photons are scattered in
the spontaneous regime while anti-Stokes signal decay exponentially
in time. The dark shadow regions correspond to the pulse sequence
for results in Fig. 4(c) of the main manuscript. Inset: laser pulses
and image intersifier pulse sequence. \label{fig:timeEvolution}}
\end{figure}

In the main body of the article (see Fig. 4) we presented 
the coincidences between the photons scattered in the write and read processes at any
angle. Due to the limitations of our detection system, the measurements of coincidences presented in the main manuscript were time integrated thus carrying no information about the arival time of a photon \citep{Eisaman2005S}. However, additional information about the timing of the photons can
be gathered by gating the image intensifier for shorter periods of time.
By scanning a short gate window of $\tau=100\:\mathrm{ns}$ duration across
the write and read pulses we measured the temporal evolution of the average
intensity of light scattered in write and read processes as depicted
in Fig. \ref{fig:timeEvolution}.

The light scattered during write and read processes is separated with interference
filters, cropped and focused on separate regions on the camera. We
verified that the write region is not illuminated during read pulse
and read region is not illuminated during write pulse. The measurement
in Fig. \ref{fig:timeEvolution} was done with write and read pulses
longer than required for a single photon level. Note almost constant
write intensity and an exponential decay $\exp(-\gamma_{0}t)$ with
$\gamma_{0}=3.82\:\mathrm{{MHz}}$ of the average read intensity. 

To scatter light at the single photon level (results presented in Fig. 4(c) of the
main manuscript) the pulse durations were shortened to $t_{\mathrm{w}}=250\:\mathrm{ns}$ and $t_{\mathrm{r}}=500\:\mathrm{ns}$ for write and read pulses, as marked
by dark shaded regions in Fig. \ref{fig:timeEvolution}. 
Virtually time independent intensity of scattered light in the write process is the signature of the spontaneous Stokes scattering regime.
After about 500 ns the stimulated regime begins, where more than one photon is scattered per mode and the number of scattered photons start to grow exponentially with drive pulse energy, due to the bosonic gain. Here a mode is defined as a portion of a light field and a corresponding spin-wave that are coupled exclusively to each other by a Raman interaction, thus their joint evolution can be described by two mode squeezing \citep{Kolodynski2012S}, with squeezing parameter proportional to the drive field amplitude and the Raman coupling strength. For the size of the driving beams used in the experiment ($\pi w^2/\lambda\gg L$), the modes can be approximated by plane waves. The number of modes $M$ can be estimated assuming atomic ensemble is a rectangular box with dimensions $w\times w\times L$.

Consider $M$-pairs of spin-wave and photon modes. The driving beam during Stokes scattering produces photon-spin-wave pairs with probability $\zeta$ in each mode. As long as $\zeta\ll 1$, the number of photons is a linear function of drive pulse energy and the total number of photons $\avg[n]=M\zeta$. Therefore, we are in the linear regime as long as the total number of photons is smaller than the number of modes: $\zeta M\ll M$. As compared with Fig. \ref{fig:timeEvolution}, we can see that indeed the photon production rate bends up when the average number of photons per angular mode becomes equal to 0.47.

Figure 4(c) depicted the situation with 0.9 Stokes--anti-Stokes photon pairs in the whole memory volume. Taking into account the read efficiency in the low-gain regime $\eta^{\mathrm{(low)}}_{\mathrm{read}}=0.13$, we generated 6.9 Stokes--spin-wave pairs during the write stage which for $M=38$ angular modes corresponds to the 0.18 generated photon--spin-wave pairs per mode. That situation can be compared with the results depicted in Fig. \ref{fig:timeEvolution} where the dark shadow regions represent the situation from Fig. 4(c). After 250 ns (which is still in the spontaneous regime) the total number of scattered photons is 0.21 per angular mode. The number of modes produced in the write process is about $M=130$. The results are corrected for finite heralding, detection and read efficiency and the fact that only c.a. 40\% of the scattered light produce spin-wave excitations \citep{Dabrowski2015S}. 

This reinforces our statement that for a pulse duration of 250 ns, on average no more than a single photon--spin-wave pair is created per mode. In turn exponential decay of the read light confirms conversion of the stored spin-waves to light \citep{Dabrowski2014S}. Because in the low-gain regime 58\% of the total number of detected pairs are accidentals, it is extremely hard to detect non-classical cross-correlations \citep{Albrecht2015S} between the photons.

We also conducted correlation measurements for pulse sequence depicted
in Fig. \ref{fig:timeEvolution} and gate pulses either: containing
whole write, or second half of write, first half or whole read pulse.
The strongest correlations are found between Stokes and anti-Stokes light scattered during whole write and read
pulses. The Stokes--anti-Stokes correlations during whole write and first half of the
read pulse are almost of the same strength. The correlations during
second half of the write pulse and the either whole or first half
of the read are notably weaker though. This is in concert with the
observation that the write operation lasts for the entire pulse, while
the read is decaying fast.

In summary, all the above evidence supports the conclusion that the
information is being transferred from the Stokes scattering to the anti-Stokes scattering
occurring afterwards.

\section{Decay of the mode number is due to diffusion}

\begin{figure}[b]
\includegraphics[scale=0.32]{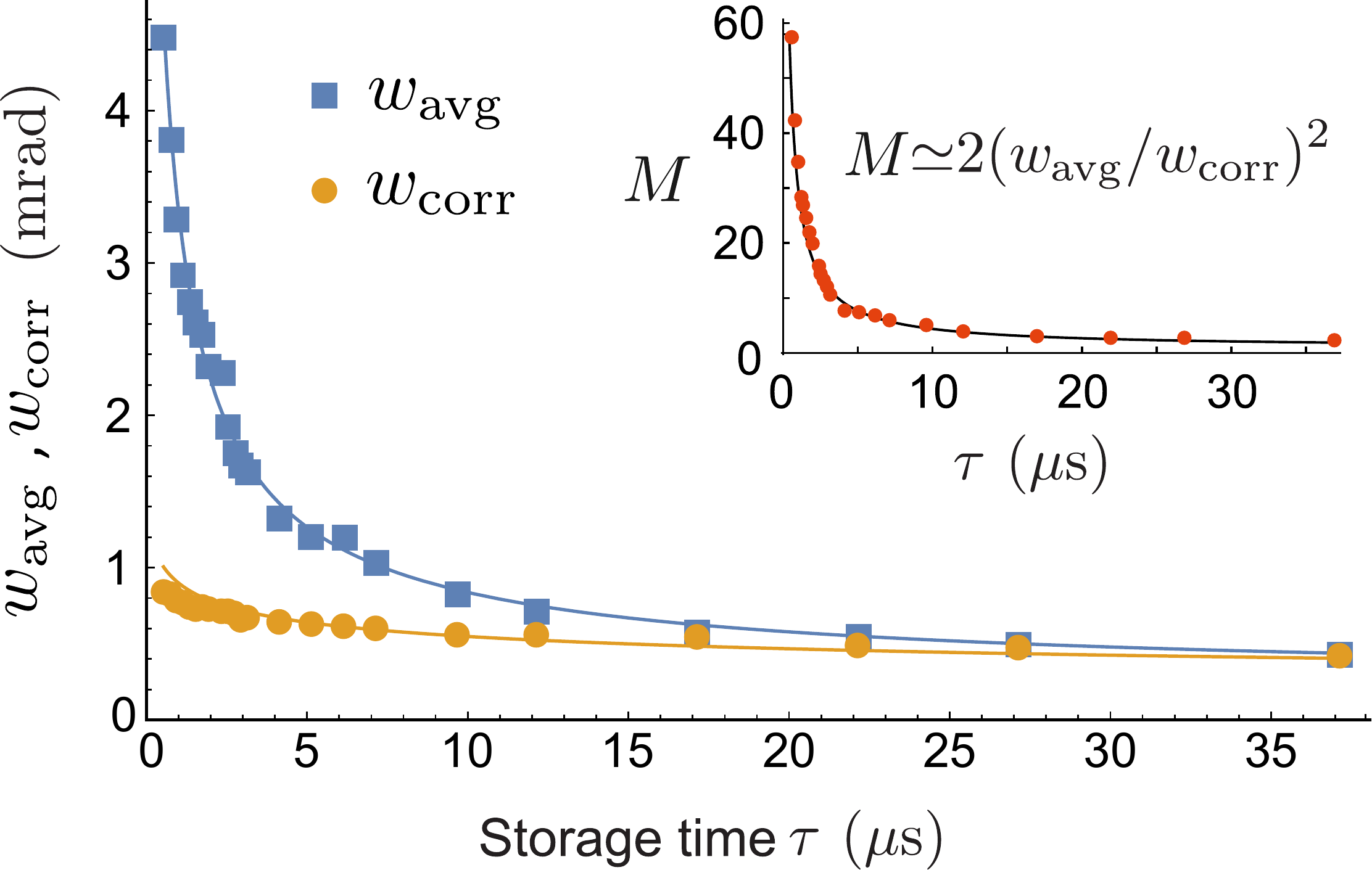}\centering 

\caption{Variances of coincidence distribution in diagonal $w_{\mathrm{avg}}$
and anti-diagonal $w_{\mathrm{corr}}$ directions versus storage time
$\tau$. Inset: estimated number od modes stored inside the memory.
Compare Fig 4(b) of the main manuscript.\label{fig:diffusion}}
\end{figure}

We quantify the angular properties of the Raman scattering by measuring
and analyzing distributions of coincidences $n_{\mathrm{coinc}}(\boldsymbol{\theta}^{(\mathrm{S})},\boldsymbol{\theta}^{(\mathrm{AS})})$
between the Stokes and time-delayed anti-Stokes photons, where for
simplicity we focus on bidimensional coincidence distributions such
as $n_{\mathrm{coinc}}(\theta_{x}^{(\mathrm{S})},\theta_{x}^{(\mathrm{AS})})$.
The correlation maps presented in Fig. 4(a) of the main manuscript are
virtually 2D gaussians with standard deviations $\sigma_{1}(t)$ and
$\sigma_{2}(t)$ of the form: 
\begin{equation}
n_{\mathrm{coinc}}(t)=n_{\mathrm{0}}\exp\left(-\frac{(\theta_{x}^{(\mathrm{S})}+\theta_{x}^{(\mathrm{AS})})^{2}}{2\sigma_{1}^{2}(t)}-\frac{(\theta_{x}^{(\mathrm{S})}-\theta_{x}^{(\mathrm{AS})})^{2}}{2\sigma_{2}^{2}(t)}\right).
\end{equation}
The diffusion attenuates the spin-waves with wave-vector $\mathbf{K}$ at a
rate $\gamma(\mathbf{K})=D\lvert\mathbf{K}\rvert^{2}$ where $D$ is the diffusion coefficient.
Due to attenutaion of spin-waves, the read photons are scattered in
narrower range of angles $\theta_{x}^{(\mathrm{AS})}$. Therefore the
number of coincidences decays with time:
\begin{equation}
n_{\mathrm{coinc}}(t)=n_{\mathrm{coinc}}(0)\exp\left(-2Dt\left(\frac{2\pi}{\lambda}\theta_{x}^{(\mathrm{AS})}\right)^{2}\right).
\end{equation}
Thus both variances $\sigma_{1}^{2}$ and $\sigma_{2}^{2}$ decay
according to $\left[\sigma_{\mathrm{i}}^{2}(t)\right]^{-1}=\left[\sigma_{\mathrm{i}}^{2}(0)\right]^{-1}+8\pi^{2}Dt/\lambda^{2},\:\mathrm{i}=1,2$.

We also take into consideration the finite angular extent (finite
spatial size) of the read laser beam diameter $2w$ which 
means the angular distribution we finally register $\tilde{n}_{\mathrm{coinc}}(\theta_{x}^{(\mathrm{S})},\theta_{x}^{(\mathrm{AS})},t)$
is described by the convolution of the atomic contribution $n_{\mathrm{coinc}}(\theta_{x}^{(\mathrm{S})},\theta_{x}^{(\mathrm{AS})},t)$
and the gaussian angular spread of the read laser pulse: 
\begin{equation}
\tilde{n}_{\mathrm{coinc}}(t)=n_{\mathrm{coinc}}(t)\ast\exp\left(\frac{-(\theta_{x}^{(\mathrm{AS})})^{2}}{w^{2}}\right).
\end{equation}
From the above formula we calculate the widths $w_{\mathrm{avg}}$
and $w_{corr}$ in diagonal and anti-diagonal directions, respectively.

The estimated number of modes $M$ retrieved from our memory can be
taken as the ratio of the solid angle span of anti-Stokes to a single
mode spread $M\simeq2(w_{\mathrm{avg}}/w_{\mathrm{corr}})^{2}$ \citep{Chrapkiewicz2012S},
taking into account full two-dimensional mode structure. Such a figure
of merit is completely equivalent with the entanglement dimension
coefficient $\mathbf{D}$ \citep{Edgar2012S}, with the correspondence
$M=2\mathbf{D}$.

The Fig. \ref{fig:diffusion} (see Fig. 4(b) of the main manuscript)
presents the decay of both widths $w_{\mathrm{avg}}$ and $w_{corr}$
with time together with theoretical curve obtained from the above
formulas. The decay of the number of modes $M$ with time is due to
diffusion, but it is not exponential.

\section{Independence of modes}

Different modes of atomic spin-wave excitations remain independent during storage time and any cross-influence of modes can be excluded. This is evidenced by the same angular spread of anti-Stokes light, conditioned on certain Stokes direction - regardless on the storage time (see Fig. 4(a) of the main manuscript).

The diffusion picture reinforces this statement. A flat plane spin-wave
with wave-vector $\mathbf{K}$ will decay due to diffusion, but its wave-vector
will not be changed and will not spread. In our experiment we create
spin-waves of 3.2 mm diameter by detecting scattered light in the far field.
Upon creation they consist of a range of wave-vectors around central
$\mathbf{K}$. Due to diffusion the initial distribution in the momentum space
will be distorted towards small $\mathbf{K}$, however will never get broader
than it was initially. Therefore two neighboring (in $\mathbf{K}$ space) modes
will never cross talk. Alternatively, we can consider the cell in the
position representation: both neighboring modes diffuse in time - their coherence
spreads over more and more volume. Nonetheless the spin-wavefronts directions
remain the same, implying that the overlap of the modes never rises.

It is worthwhile to consider how the atoms carry the information contained
in the spin-waves. In our scheme any atom is potentially carrying information
about each illuminated mode. This situation can be opossed to \citep{Lan2009S}
where the atomic cloud was divided into 12 memories one next to
the other. In that experiment, where spatial, not angular modes were used, the total number of atoms was practically
split to form twelve separate ensembles. It is noteworthy that with the configuration from \citep{Lan2009S}, after very long storage times the atoms from one spatial
region may diffuse to neighboring spatial regions.

With the same thermal speed of atoms or diffusion coefficient $D$ both methods (our and that presented in \citep{Lan2009S})
give the same OD for the same number of modes for the same storage
time. In our method the ensemble diameter is divided by the number
of modes to get the maximal distance the atoms can travel before the
highest mode is washed out. If multiple separate ensembles are used,
the diameter of each has to be smaller than this distance. Thus the atoms
have to fill the same cross section area, leading to the same OD.

\end{document}